\documentclass[aps,amsmath,citeautoscript,longbibliography,nofootinbib]{revtex4-2}
\usepackage[utf8]{inputenc}
\usepackage{bm}
\usepackage{tensor}
\usepackage[italicdiff]{physics}
\usepackage{amsmath,mathrsfs,amsfonts}
\usepackage{graphicx,afterpage}
\usepackage{subcaption}
\usepackage{amsmath,latexsym}
\usepackage{color,soul}
\usepackage{float}
\usepackage[section]{placeins}
\usepackage{silence}
\usepackage[hidelinks]{hyperref}
\usepackage{lipsum}
\begin{document}
\title{New Lorentzian Taub-NUT and Euclidean Eguchi-Hanson Solutions in $f(R)$ gravity}
\author{Joshua G. Fenwick}
\email{joshua.fenwick@usask.ca}
\author{Masoud Ghezelbash}
\email{masoud.ghezelbash@usask.ca}
\affiliation{Department of Physics and Engineering Physics, University of Saskatchewan, Saskatoon SK S7N 5E2, Canada}
\date{\today}
\begin{abstract}
A New Taub-NUT black hole in multiple coordinate systems is solved and analyzed. This black-hole is found to be an extension of the Clifton-Barrow metric for the specific case that $\delta=-1/2$. We look at physical properties such as test particle orbits, thermodynamics, and various space-time limits. In addition, we show an Eguchi-Hanson type-II $f(R)$ solution related to the found Taub-NUT framework.
\end{abstract}
\maketitle
\section{Introduction}
The gravitational instantons are the most interesting class of exact solutions to the Einstein equations. The gravitational instantons are regular and complete solutions everywhere. These solutions posses a self dual curvature two form.  The gravitational instantons exist in vacuum \cite{P1} or exist in presence of a cosmological constant \cite{P2}. Among these solutions, Taub-NUT solutions and its bolt extension, are the most well known solutions \cite{P3}. The Taub-NUT solutions have a lot of applications,  as in higher-dimensional gravity \cite{P17}, M-theory \cite{P5}, black hole holography \cite{P6} and black hole thermodynamics \cite{P7}. Another interesting class of the gravitational instantons are Eguchi-Hanson spaces \cite{P8} and Atiyah-Hitchin spaces \cite{P9}. The Eguchi-Hanson and Atiyah-Hitchin spaces have been used extensively in construction of higher-dimensional solutions to extended theories of gravity \cite{P10,P11,P12}, supergravity \cite{P13}, \cite{P14}, as well as in investigating the quantum properties of the black holes \cite{P15}.
Moreover, the gravitational instantons are the dominant metrics in the path-integral formulation of the Euclidean quantum gravity. These dominant metrics are also related to the minimal surfaces in Euclidean space \cite{P16}. In fact, the two dimensional minimal surfaces provide solutions to the real elliptic Monge-Amp\`ere equation on a real two dimensional manifold. The solutions are the base for the Kahler metrics for some gravitational instantons beside the aforementioned well known solutions \cite{P17}.
Inspired by the existence of the gravitational instantons in four dimensions, in this article, we find exact instanton solutions in $f(R)$ gravity.
The $f(R)$ gravity is the simplest version of 
the modified theories of gravity (MTG) \cite{B23,B22,B26,B28,B29,B42}, in which the Ricci scalar $R$ in the Einstein-Hilbert action is replaced by a function $f$ of $R$.  The MTG
are theories to address the issues of GR, such as dark energy since the accelerating expansion
of the universe was discovered. 

We also note that in the appropriate limits, the MTG should
produce the results that are in agreement with GR. In the modified theory of $f(R)$ gravity,
the action includes a function of Ricci scalar $R$, which leads to a new class of MTG.
Other theories of MTG include $f(T)$ and $f(Q)$ gravity. In the former case, the Ricci scalar $R$ is replaced by the torsion scalar $T$ in the Einstein-Hilbert action (and the theory is called Teleparallel Equivalent of General Relativity (TEGR)) and then the TEGR theory is upgraded to a function $f$ of $T$ \cite{Spherical,43,44,45,46,47,48,50,51}.
In the latter case, we consider the non-metricity scalar Q, instead of Ricci or Torsion scalars in the action of general relativity (The theory is called Symmetric Tele-parallel Equivalent of General Relativity (STEGR)), and then upgrade the theory to a function $f$ of $Q$. We should note that TEGR and STEGR are equivalent to general relativity. However considering a more general theory, where the action is a function of non-metricity scalar $f(Q)$, leads to a new class of MTG \cite{52,53,54,55,56}. Moreover, we can consider other more general theories of gravity, where the action is a function of more than one of the $R$, $T$, $Q$ or other scalars. As an example, $f(R,\mathcal{T})$ theories with $\mathcal{T}$ the stress-energy scalar were considered in cosmological model building and observational constraint \cite{57,58,59,60,61,62,63,64,65,66}. 

In this article, we construct the Taub-NUT and Eguchi-Hanson solutions to a specific $f(R)$ gravity and study its behaviours. 
{\textcolor{black}{The article is the first step in constructing the rotating solutions with the NUT twist in the context of $f(R)$ gravity. Once constructed, it is an open question to establish or rule out the existence of the conformal symmetry for the rotating NUT solutions in $f(T)$ gravity \cite{BMe}.
}}
The paper is organizes as follows: In section \ref{sec1}, we review the $f(R)$ gravity and the field equations. In section \ref{sec2}, we consider the Taub-NUT solutions in $f(R)$ gravity, and find two classes of solutions. We discuss the physical properties of the solutions. 
In section \ref{sec3} we consider the Eguchi-Hanson solutions in $f(R)$ gravity, and discuss its physical properties.

\section{Metric $f(R)$ gravity}\label{sec1}
Metric affine $f(R)$ is one of the methods of deriving a formulation of $f(R)$, with the others being Palatini and metric-affine $f(R)$ \cite{Palatini_f(R),Lovelock}. In metric $f(R)$ gravity we assume, a priori, that our connection is defined to be the Levi-Civita connection. With this formulism our action takes the form \cite{Palatini_f(R),Lovelock}
\begin{equation}
    \mathcal{S}_{f(R)} = \dfrac{1}{2\kappa}\int_\mathcal{M}\dd[4]{x}\sqrt{||g||}f(R) +\dfrac{\varepsilon}{\kappa}\int_{\partial\mathcal{M}}\dd[3]{x}\sqrt{||h||}f_R \mathcal{K}+ \mathcal{S}_M\:.\label{METRICFOFR}
\end{equation}
We will not need make use of the extended Gibbons-Hawking-York boundary term but we include it for completeness. Notation wise, we denote the derivatives with the Ricci as subscripts such as $df/dR \equiv f_R$ and save prime notation to solely define derivatives with respect to the coordinate $r$. Taking the variation with respect to the metric results in the standard $f(R)$ field equations as 
\begin{equation}
    f_RR_{\mu\nu} - \dfrac{1}{2}f(R)g_{\mu\nu}-\qty[\nabla_\mu\nabla_\nu-g_{\mu\nu}\Box]f_R = \kappa\mathcal{T}_{\mu\nu}\:,\label{FR_met_EOM}
\end{equation}
with a trace given by 
\begin{equation}
    f_R R-2f(R)+3\Box f_R =\kappa \mathcal{T}\:.
\end{equation}
Here, $\Box$ is the standard GR definition of the d'Alembertian as $g_{\mu\nu}\nabla^\mu\nabla^\nu = \Box$. 
Alteration of the field equations one finds the alternate expression
\begin{equation}
    G_{\mu\nu} = \dfrac{\kappa}{f_R}\mathcal{T}_{\mu\nu}-\dfrac{1}{2f_R} \qty(f_RR-f(R))g_{\mu\nu} +\dfrac{1}{f_R}\qty[\nabla_\mu\nabla_\nu-g_{\mu\nu}\Box]f_R\:,\label{Met_fr_BD}
\end{equation}
where we can identify metric $f(R)$ gravity to take the form of a Brans-Dicke theory with $\phi=f_R$ and $\omega_0=0$ \cite{Brans_Dicke}.

\section{Taub-NUT extension to the $\delta=-1/2$ Clifton–Barrow case}\label{sec2}
We set out to examine the Taub-NUT \cite{Taub,NUT} charge within the framework of the modified theories of gravity, for that reason we begin with a metric ansatz 
\begin{equation}
    ds^2=-B(r)e^{2\alpha(r)}\qty(dt+2n\cos(\theta)d\varphi)^2+\dfrac{dr^2}{B(r)}+(r^2+n^2)d\Omega^2\:,
\end{equation}
and hope to solve for $\alpha(r)$, $B(r)$, and $f(R)$. To do so we extend methods of \cite{FR_TN_METHOD} to include the NUT charge. We express our action as
\begin{equation}
    \mathcal{S}=\int\dd[4]{x}\sqrt{-g}\qty(f(R)-f_R(R-R_{ex}))=\int\dd[4]{x}\sqrt{-g}\mathcal{L}\:,
\end{equation}
with $R_{ex}$ defined to be the exact Ricci expression calculated from our ansatz. Simplification of our action results in the expression 
\begin{align}
\sqrt{-g}\mathcal{L}&=e^{\alpha}\qty(r^2+n^2)\left[f(R)-f_R\left(R+2B\qty(\dv{\alpha}{r})^2 + 2B\dv[2]{\alpha}{r} + \dv[2]{B}{r} -\dfrac{4r}{r^2+n^2}\dv{B}{r}\right.\right.\nonumber\\&-\left.\left.\qty(3\dv{B}{r}+\dfrac{4rB}{r^2+n^2})\dv{\alpha}{r}+\dfrac{2n^2Be^{2\alpha} + (-4n^2 - 2r^2)B}{(n^2 + r^2)^2}+\dfrac{2}{r^2+n^2}\right)\right]\:.\label{Lag_no_sin}
\end{align}
Before we vary this action we make use of integration by part to remove second derivatives of the metric functions a receive the expression
\begin{align}
    \sqrt{-g}\mathcal{L} &= e^{\alpha}\left[\qty(r^2+n^2)\qty(f(R)-f_RR) + 2f_R\qty(1-\dv{B}{r}r-B+\dfrac{Bn^2\qty(e^{2\alpha}-1)}{r^2+n^2})\right.\nonumber\\&+\left.f''(R)\dv{R}{r}\qty(r^2+n^2)\qty(2B\dv{\alpha}{r}+\dv{B}{r})\right]\:.
\end{align}
Our method now is to vary this action with respect to $\alpha$ and $B$ as opposed to the metric components $g_{\mu\nu}$. This methodology will result in 2 second order equations as opposed to the fourth order expressions in (\ref{FR_met_EOM}). Using the Euler-Lagrange equations, we may find the variations. Variation with respect to the function $\alpha(r)$ leads us to find  
\begin{align}
    \fdv{\sqrt{-g}\mathcal{L}}{\alpha} &= \dfrac{-e^{\alpha}}{r^2+n^2}\left[n^2\qty(f_R\qty(4B-6e^{2\alpha}B-2)+n^2\qty(2\dv[2]{f_R}{r}B+\dv{B}{r}\dv{f_R}{r}+f_RR-f(R)))\right.\nonumber\\&+2n^2 r\qty(2B\dv{f_R}{r}+f_R\dv{B}{r})+2r^2\qty(n^2\qty(2\dv[2]{f_R}{r}B+\dv{B}{r}\dv{f_R}{r}+f_R R-f(R))+f_R\qty(B-1))\nonumber\\ &+\left.2r^3\qty(2B\dv{f_R}{r}+f_R\dv{B}{r}) +r^4\qty(2\dv[2]{f_R}{r}B+\dv{f_R}{r}\dv{B}{r}+f_R R-f(R))\right]\:.\label{Varalpha}
\end{align}
Variation with respect to $B(r)$ is simpler and leads us to find 
\begin{align}
    \fdv{\sqrt{-g}\mathcal{L} }{B} &= \dfrac{e^{\alpha}}{r^2+n^2}\left[n^4\qty(\dv{f_R}{r}\dv{\alpha}{r}-\dv[2]{f_R}{r})+2n^2\left(f_R\qty(2e^{2\alpha} +2r\dv{\alpha}{r}-2)\right.\right.\nonumber\\&+\left.\left.r^2\qty(\dv{\alpha}{r}\dv{f_R}{r}-\dv[2]{f_R}{r})\right)-r^3\qty(r\qty(\dv[2]{f_R}{r}-\dv{\alpha}{r}\dv{f_R}{r})-2f_R\dv{\alpha}{r})\right]\:.\label{VarB}
\end{align}
Equations (\ref{Varalpha}) and (\ref{VarB}) will be the two expression we solve to find our Taub-NUT spacetime. Given we now have 2 expressions to solve for $\alpha(r)$, $B(r)$, and $f(R)$ we will institute a guess and check approach for $\alpha(r)$.
\subsection{Solution set I}
We begin with the naive guess that with $\alpha(r)=0$ we may find a standard Taub-NUT ansatz solution. With this we find equation (\ref{VarB}) reduce to tell us 
\begin{equation}
    \dv[2]{f_R}{r} = 0 \Rightarrow f_R = C_1r + C_2\:.\label{f_R 1}
\end{equation}
With $f_R$ and $\alpha(r)$ we will solve equation (\ref{Varalpha}) for $B(r)$. subbing in $f_R$ and taking the derivative of (\ref{Varalpha}) to replace $f(R)$ with $f_R\dv{R}{r}$ we find the differential equation

\begin{equation}
    (n^2 + r^2)^2(C_1 r + C_2)\dv[2]{B}{r} + C_1(n^2 + r^2)^2\dv{B}{r} + (4C_1 n^2 r - 4C_1 r^3 + 6C_2n^2 - 2C_2r^2)B + 2(n^2 + r^2)(C_1 r + C_2) =0\:,
\end{equation}
for which we find no closed form solution. If we remove the GR limit term $C_2$ by setting $C_2=0$ we can find the differential equation reduce and find the expression 
\begin{equation}
    B(r) = \dfrac{n^2\qty(3-2\Lambda)\ln(r^2/n^2) -r^4\Lambda - 4r^2\Lambda n^2 + 3r^2 - n^4\Lambda - \beta n^2 - 6C}{6n^2+r^2}\:.
\end{equation}
This function results in the Ricci scalar
\begin{equation}
    R = \dfrac{3 + 6\Lambda r^2 - 2\Lambda n^2}{3r^2}\:,
\end{equation}
for which we may invert to find 
\begin{equation}
    r = \sqrt{\dfrac{2\Lambda n^2 -3}{3\qty(2\Lambda - R)}}\:.
\end{equation}
With this inversion and equation (\ref{f_R 1}) we may substitute and integrate to find our $f(R)$ form to be given as 
\begin{equation}
    f(R) = \eta\sqrt{R-2\Lambda}\:,
\end{equation}
where $\eta$ is a redefinition of $C_1$ absorbing redundant constants. 
\subsection{Solution set II}
In our second solution set we take a separate way to determine our function $\alpha(r)$ We notice in (\ref{VarB}) if we take all coefficients of terms with $f(R)$ we find the equation  
\begin{equation}
    n^2\qty(2e^{2\alpha} + 2r\dv{\alpha}{r}-2) + 2r^3\dv{\alpha}{r}=0\:,
\end{equation}
which when solved gives 
\begin{equation}
    \alpha(r) = \dfrac{1}{2}\ln(\dfrac{r^2}{r^2+\epsilon(r^2+n^2)})\:.
\end{equation}
From this solution we see $\epsilon=0$ reduces to the aforementioned solution set. With this we may solve equation (\ref{VarB}) to be
\begin{equation}
    r((1 + \epsilon)r^2 + \epsilon n^2)\dv[2]{f_R}{r} - \epsilon\dv{f_R}{r}n^2 =0 \Rightarrow f_R = C_1\sqrt{r^2+\epsilon\qty(r^2+n^2)} +C_2\:. \label{f_R 2}
\end{equation}
The resultant equation for $B(r)$ becomes much more cumbersome with the non zero $\epsilon$. We do not show the equation but show that the solution is given by 
\begin{align}
    B(r) &= \dfrac{r^2+\epsilon(r^2+n^2)}{6r^2(r^2+n^2)(1+\epsilon)^2}\left( n^2\qty(3-\dfrac{2\Lambda n^2}{\epsilon+1})\ln(\dfrac{\epsilon(r^2+n^2)+r^2}{n^2}) - \Lambda(1 + \epsilon)r^4 - 2\Lambda(\epsilon + 2)r^2n^2\right.\nonumber\\ &+ \left.3(1 + \epsilon)r^2 - \Lambda(1 + \epsilon)n^4 - (1 + \epsilon)\beta n^2 - 6(1 + \epsilon)C \right)\:,
\end{align}
which leads to the Ricci scalar 
\begin{equation}
    R = \dfrac{3(1 + \epsilon) + 6\Lambda r^2(1 + \epsilon) + 2\Lambda n^2(3\epsilon - 1)}{3r^2+3\epsilon\qty(r^2+n^2)}\:.
\end{equation}
This Ricci also allows for a inversion to the form
\begin{equation}
    r = \sqrt{\dfrac{n^2\qty(3\epsilon R + 2\Lambda(1-3\epsilon))-3(1+\epsilon)}{3\qty(2\Lambda(1+\epsilon)-R(1+\epsilon))}}\:,
\end{equation}
which again we find
\begin{equation}
    f(R) = \eta\sqrt{R-2\Lambda}\:.
\end{equation}
Again, we set $\eta$ by absorbing arbitrary constant to simplify the $f(R)$ form.
\subsection{Physical properties}
We will analyze a number of physical interpretations of this extended Taub-NUT solution. We will look at a coordinate transform that links the two solutions, followed by limits of the solution and reduction to the Clifton-Barrow metric case. In addition, we will examine thermodynamics and properties of orbits when within the constraint that $\Lambda=0$.  
\subsubsection{Equality of solutions for arbitrary epsilon}
Before we analyze all other physical properties we will show how both solutions can be connected without the limit as $\epsilon=0$. This shows while $\epsilon$ may appear as a functional generalization it can be systematically removed from the solution.
Starting first from solution set II we may perform the coordinate transform
\begin{equation}
    r\Rightarrow \sqrt{r^2-\epsilon m^2}\:,
\end{equation}
and
\begin{equation}
       t \Rightarrow t\sqrt{\epsilon+1}\:.
\end{equation}
In addition we define $m$ as a new NUT parameter related to the old via
\begin{equation}
    n \Rightarrow m\sqrt{\epsilon+1}\:.
\end{equation}
Lastly we simplify some term coefficients and the arbitrary constant $C$ to remove extra $\epsilon$ dependence as in the replacement 
\begin{equation}
    \beta \Rightarrow \dfrac{\beta-3\epsilon}{1+\epsilon}\:,
\end{equation}
and
\begin{equation}
    C \Rightarrow  \dfrac{\Lambda \epsilon m^4}{3} + C\:.
\end{equation}
Following these redefinitions, we arrive back to the solution given by solution set I. With the result that both black-holes are physically equivalent for any $\epsilon$ we could perform all analysis subject to $\epsilon=0$ to achieve the same physical ideals. For the sake of completion we will analyze in the more ``general" spacetime leaving $\epsilon$ intact as the substitution $\epsilon=0$ limit reduces to the simpler case.

\subsubsection{Space-time singularities and limits}
To discuss the singularities of the found solution of the spacetime we can analyze the denominator of the Kretschmann invariant. Calculation of the Kretschmann, we find it to be  
\begin{equation}
    K = \dfrac{\frak{K}(r)}{\qty(1+\epsilon)^6\qty(n^2+r^2)^6\qty[r^2+\epsilon(n^2+r^2)]^2}\:.
\end{equation}
We see a problem for for $\epsilon=-1$ which results from equation (\ref{f_R 2}) losing $r$ dependence. Further we see a singularity at $r=n=0$ just as exists in the standard GR Taub-NUT. The final possible singularity exists at the value 
\begin{equation}
    r = \pm n\sqrt{\dfrac{-\epsilon}{1+\epsilon}}\:,
\end{equation}
which is only a real singularity given $\epsilon$ falls in $-1<\epsilon<0$. However, we can constrain allowed $\epsilon$ values when looking at the limits of our spacetime. Should we analyze the limits should $r$ approach infinity, we find this limit approach to the metric
\begin{equation}
    g_{\mu\nu} = \begin{pmatrix}
        \dfrac{\Lambda\epsilon}{6\qty(1+\epsilon)}r^2 & 0 & 0 & \dfrac{n\cos(\theta)\Lambda\epsilon}{3\qty(1+\epsilon)}r^2\\
        0 & -\dfrac{6}{\Lambda r^2} & 0 & 0 \\
        0 & 0 & r^2 &  0 \\
        \dfrac{n\cos(\theta)\Lambda\epsilon}{3\qty(1+\epsilon)}r^2 & 0 &  0 & r^2\qty(\sin^2\theta+\dfrac{2n^2\cos[2](\theta)\Lambda\epsilon}{3\qty(1+\epsilon)})
    \end{pmatrix}\:,
\end{equation}
while if we let $\Lambda=0$ we find the limit
\begin{equation}
    g_{\mu\nu} = \begin{pmatrix}
        -\dfrac{\epsilon}{2\qty(1+\epsilon)} & 0 & 0 & -\dfrac{n\epsilon\cos(\theta)}{1+\epsilon}\\
        0 & 2 & 0 & 0 \\
        0 & 0 & r^2 &  0 \\
        -\dfrac{n\epsilon\cos(\theta)}{1+\epsilon} & 0 &  0 & r^2\sin^2\theta
    \end{pmatrix}\:.
\end{equation}
From the $\Lambda=0$ limit we can see for $\epsilon<-1$ that time approaches a positive value for large $r$ and so we mandate that $\epsilon>-1$ must hold to preserve the Lorentzian structure. Additionally, if we examine the limit that $n=0$ we can find $\alpha(r)$ approach 
\begin{equation}
    \alpha(r) = \dfrac{1}{2}\ln(\dfrac{1}{1+\epsilon})\:,
\end{equation}
while $B(r)$ approaches 
\begin{equation}
	B(r)=\dfrac{1}{2} - \dfrac{C}{r^2} - \dfrac{\Lambda}{6}r^2\:.
\end{equation}
$\alpha(r)$ can be removed via a redefinition of time and we see the resultant metric is the $\delta=-1/2$ case of the Clifton-Barrow solution \cite{clifton2005power} plus a cosmological constant. Thus we have found a CB extension into the Taub-NUT regime for a specific case of Ricci power.
\subsubsection{Black hole thermodynamics}
To analyze the thermodynamics, we will derive the expressions of our space-time. Following \cite{Thermo_G,thermo_fR}, we begin by equating the field equations at the horizon point $r_+$ shown by
\begin{equation}
    \eval{8\pi\tensor{\mathcal{T}}{^{r}_{r}}}_{r=r_+} = \eval{f_R{R_{r}^r} - \dfrac{1}{2}f(R)-\qty[\nabla_r \nabla^r-\Box]f_R}_{r=r_+}\:,
\end{equation}
where we may denote $P=\tensor{\mathcal{T}}{^r_r}$ to be the pressure \cite{Thermo_G,thermo_fR}. With our expression for $P$, we may equate it to 
\begin{equation}
    P = D(r_\pm)+C(r_\pm)T\:,
\end{equation}
to find our functions $D(r_\pm)$ and $C(r_\pm)$ with our temperature $T$ given by the usual $T=B'(r_\pm)/4\pi$. These functions take the forms
\begin{align}
    C(r_\pm) &= \dfrac{1}{4}\qty(\dfrac{2r_\pm f'(R)}{r_\pm^2+n^2} + \eval{\dv{f'(R)}{r}}_{r=r_\pm})\:,\\
    D(r_\pm) &= -\dfrac{1}{8\pi}\qty(\dfrac{f'(R)}{r_\pm^2+n^2} + \dfrac{1}{2}\qty(f(R)-Rf'(R)))\:.
\end{align}
Finally, we may take our metric determinant and integrate it over all coordinates while keeping time constant to find the volume of the black hole to be
\begin{equation}
    V(r_\pm) = \eval{\int\dd[3]{x}\sin[2](\theta)\qty(n^2+r^2)}_{r=r_\pm} = 4\pi r_+\qty(n^2+\dfrac{r_\pm^2}{3})\:.
\end{equation}
With functions $V'(r),C(r),$ and $D(r)$, we may find the temperature, entropy, and energy \cite{Thermo_G,thermo_fR}. In the case of $\Lambda=0$ we find our black-hole horizon to be given by the expression
\begin{equation}
    r_\pm = \pm n \sqrt{\dfrac{W_0(e^{\beta\qty(1+\epsilon)/3}e^{2C\qty(1+\epsilon)/n^2}e^{\epsilon})-\epsilon}{1+\epsilon}}\:,
\end{equation}
which is guaranteed to exist for strictly positive constants. Should one want negative $C$ or $\beta$ values the root may not be guaranteed to exist. To mandate a root exist we can enforce the constant constraint that  
\begin{equation}
    \epsilon \leq -\dfrac{3n^2}{6C+\beta n^2}W_0\qty(-e^{\beta/3+2C/n^2}\qty(\dfrac{\beta}{3}+\dfrac{2C}{n^2}))\:,
\end{equation}
so long as the Lambert W function is real. If the above Lambert W function is imaginary a root is guaranteed and if not then the constraint on $\epsilon$ is imposed. With this root the thermodynamic quantities can be calculated and expressed via
\begin{gather}
    T = \eval{\dfrac{1}{4\pi}\dv{B(r)}{r}}_{r=r_\pm} = \dfrac{1}{4\pi r_\pm}\:,\\
    S = \eval{\int \dd{r} V'(r)C(r)}_{r=r_\pm}= \dfrac{\eta\pi}{\qty(\epsilon+1)^3/2}\sqrt{r_\pm^2+\epsilon(r_\pm^2+n^2)}\qty(r_\pm^2+2n^2+\epsilon(r_\pm^2+n^2))\:,\\
    E = \eval{-\int \dd{r} V'(r)D(r)}_{r=r_\pm} = \dfrac{3\eta}{16\qty(\epsilon+1)^{3/2}}\qty(n^2\dfrac{3\epsilon+2}{3\sqrt{1+\epsilon}}\ln(\sqrt{1+\epsilon}\dfrac{r_\pm}{n}+\dfrac{\sqrt{r_\pm^2+\epsilon(r_\pm^2+n^2)}}{n}) + r_\pm\sqrt{r_\pm^2+\epsilon\qty(r_\pm^2+n^2)})\:.
\end{gather}
Aside from this thermodynamic analysis we reserve a more in-depth $f(R)$ Taub-NUT thermodynamic analysis for a separate publication.
\subsubsection{Test particle orbits}
Examining test particle orbits we find the radial and polar trajectory equations to be given by \cite{Carterconst}
\begin{gather}
     \qty(\dv{r}{\lambda})^2 +\dfrac{B(r)}{r^2+n^2}\qty(Q+\mu^2r^2)-\dfrac{E^2}{e^{2\alpha(r)}}= 0\:,
\end{gather}
and
\begin{equation}
    (r^2+n^2)^2\qty(\dv{\theta}{\lambda})^2 +\dfrac{\qty(L+2n\cos(\theta)E)^2}{\sin[2](\theta)}= Q -\mu^2 n^2\:.\label{FR_polar orbit}
\end{equation}
with solutions to these one can find the azimuthal and temporal equations to be 
\begin{align}
    \dv{\varphi}{\lambda} &= \dfrac{L+2n\cos(\theta)E}{\qty(r^2+n^2)\sin[2](\theta)}\:,\\
    \dv{t}{\lambda} &= \dfrac{E}{B(r)e^{2\alpha(r)}}+\dfrac{2n\cos(\theta)}{\qty(n^2+r^2)\sin[2](\theta)}\qty[L+2n\cos(\theta)E]\:.
\end{align}
These equations cannot be solved analytically in a closed form manner however, from the equations we can get some details related to the trajectories. Examining the angular equation and attempting to enforce a equatorial or a single theta orbit we can find theta must take one of the values 
\begin{equation}
    \theta = \pi -\cos[-1](\dfrac{L}{2nE})\quad\text{or}\quad\theta = \pi-\cos[-1](\dfrac{2n E}{L})\:.\label{THETAS}
\end{equation}
In addition, we must enforce the additional constraint on the Carter constant for each of the the theta choices given by 
\begin{equation}
    Q=\mu^2n^2 \quad\text{or}\quad Q=L^2+\mu^2n^2 -4E^2n^2\:.
\end{equation}
We can see from equation (\ref{THETAS}) that should $L = \pm 2nE$ we can see these theta values reduce to being the poles of the space-time. Since we have singularities on either of the space-time poles form the Taub-NUT ansatz we wish to avoid these values.
Similarly we see that if we wish to enforce $\theta = \pi/2$ we must mandate that either $L$, $E$, or $n$ equal 0 but note that the most natural is the enforcement $L=0$. Additionally, Should we enforce 
\begin{equation}
    Q \geq \mu^2 n^2+L^2\:,
\end{equation}
we can find that $\theta=\pi/2$ is included in the particles trajectory. Lastly we note that we require the constraint that 
\begin{equation}
    -1<\dfrac{L}{2nE}<1\:,
\end{equation}
or we find no theta values that would allow valid trajectories. Turning to the radial coordinate equation we cannot do much as the potential is very complex. Looking for circular orbits we can mandate that
\begin{equation}
    \dv{r}\qty(\dfrac{B(r)}{r^2+n^2}\qty(Q+\mu^2r^2)-\dfrac{E^2}{e^{2\alpha(r)}}) = 0 \label{FIRST DERIVATIVE}
\end{equation}
as well as 
\begin{equation}
    0> \dv[2]{r}\qty(\dfrac{B(r)}{r^2+n^2}\qty(Q+\mu^2r^2)-\dfrac{E^2}{e^{2\alpha(r)}}) 
\end{equation}
Solving (\ref{FIRST DERIVATIVE}) for the constant $C$ as we cannot analytically solve for $r$ we can find the constraint to become 
\begin{equation}
0 >\dfrac{-2\mu^2n^2r^2 + 2Q\mu n^2 - 4Q^2}{(-\mu n^2 + \mu r^2 + 2Q)(n^2 + r^2)r^2}\:.
\end{equation}
\begin{figure}[h!]
    \centering
    \subfloat[Plot of the potential for the parameters: $n=0.2$, $C=0.2$, $\epsilon=0.5$, $\beta=3$, $\mu=0.7$, $Q=2$, $E=0.6$\label{fig: Wikipidea a}]{\includegraphics[scale=0.4]{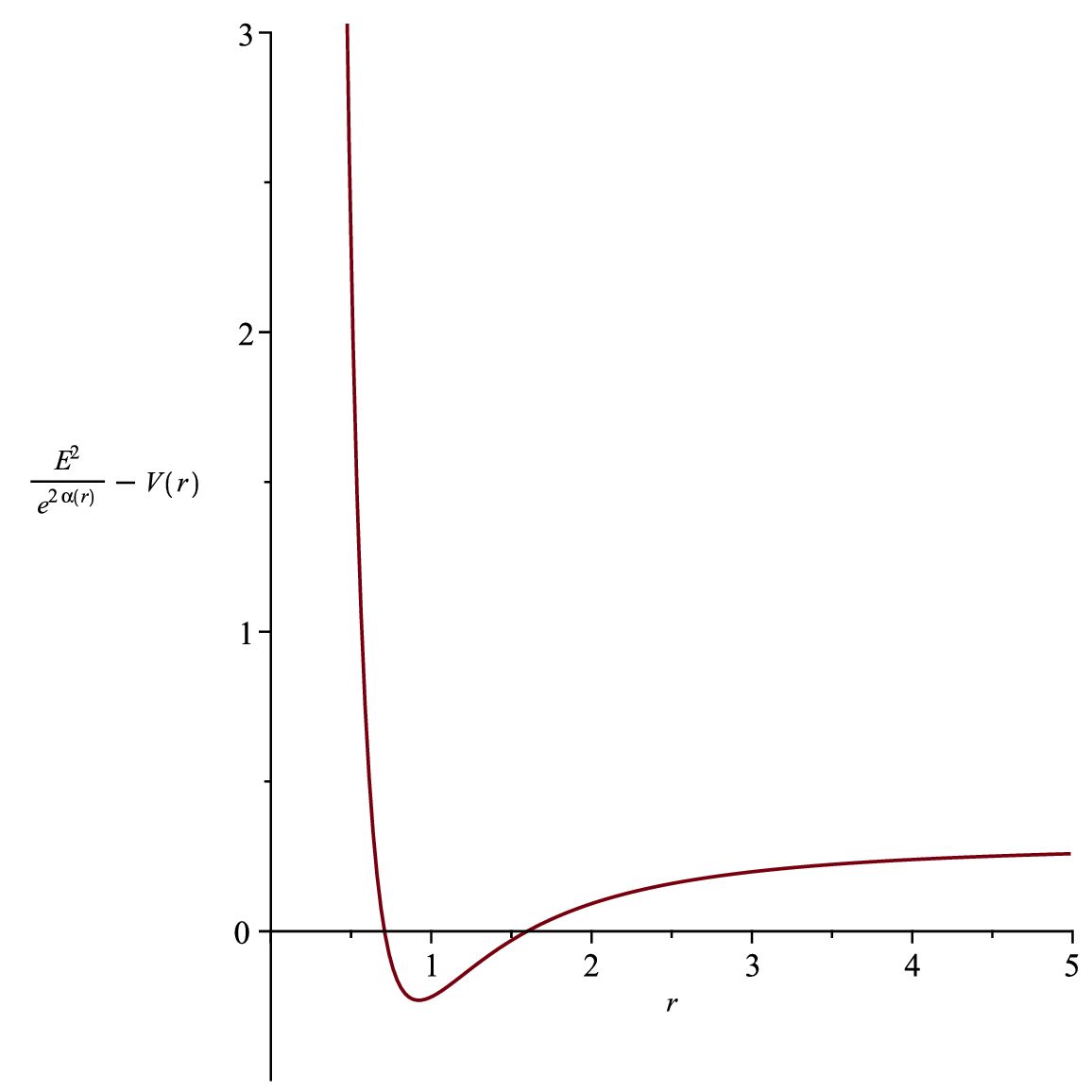}}\hfill
    \subfloat[Plot of the potential for the parameters: $n=0.2$, $C=0.2$, $\epsilon=0.5$, $\beta=10$, $\mu=1.6$, $Q=1$, $E=1.1$ \label{fig: Wikipidea b}]{\includegraphics[scale=0.4]{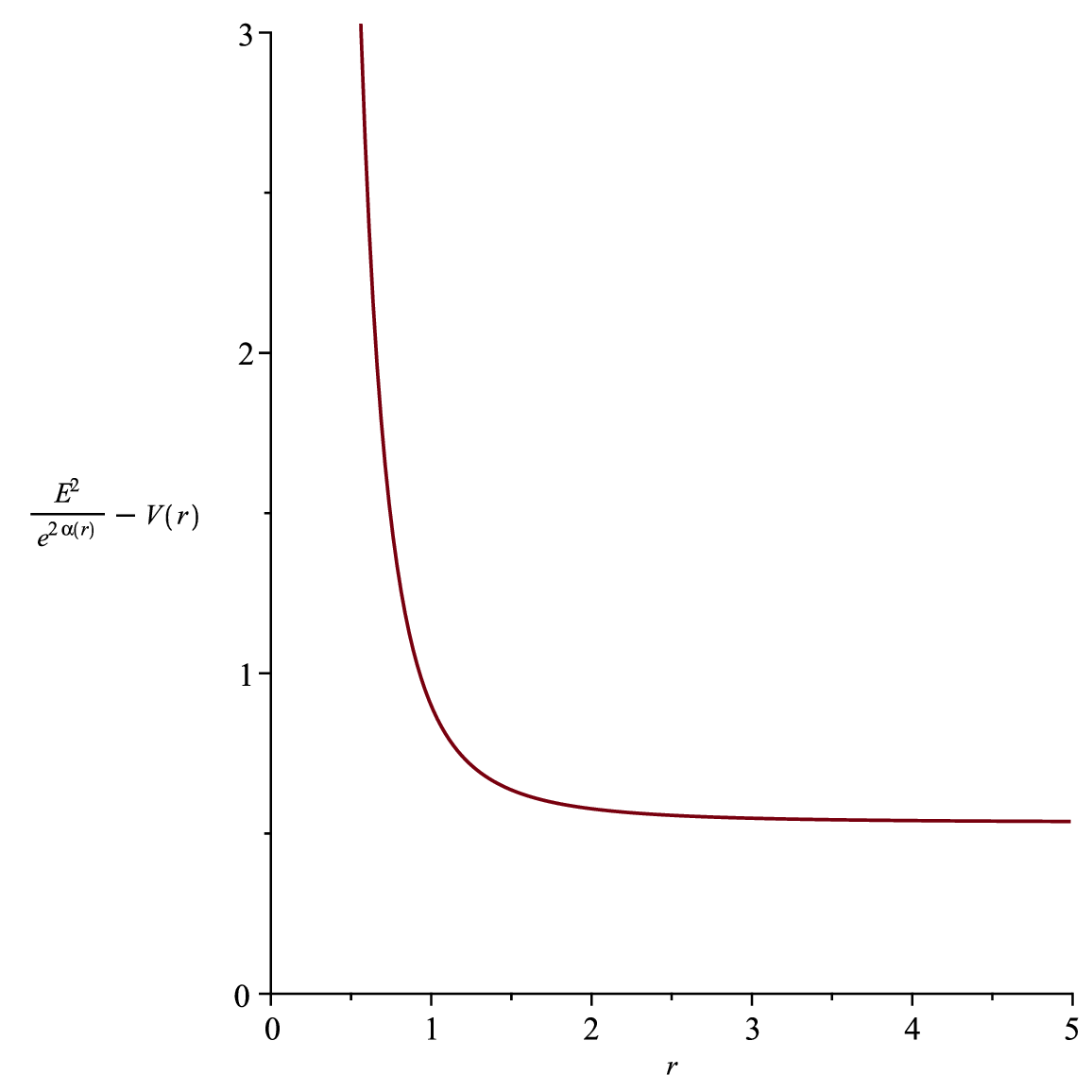}}
    \label{fig:Wikipidea}
    \caption[Plots of the test particle's orbit potential.]{Two plots of $E^2-V(r)$ for different sets of parameters. In subplot (a) we see an unstable orbit and regions in which particles are reflected or forced to fall in. In subplot (b) we see all particles are forced to fall into the black hole.}
\end{figure}
\section{An Eguchi-Hanson type-II extension}\label{sec3}
In addition to the found Taub-NUT solution we can extend this solution to a Eguchi-Hanson like space through a unique coordinate change. If we change $r$ as in 
\begin{equation}
    r \Rightarrow \dfrac{1}{2}\sqrt{r^2-\dfrac{1+\epsilon}{\xi^2}}\:,
\end{equation}
and set the time coordinate to be related to the Eguchi-Hanson angle $\psi$ via
\begin{equation}
    t = \psi\dfrac{\sqrt{1+\epsilon}}{\xi}\:.
\end{equation}
The parameter $\xi$ is related to the NUT charge as in 
\begin{equation}
    n = \dfrac{\sqrt{1+\epsilon}}{2\xi}\:,
\end{equation}
and the constant $C$ is completely redefined to be
\begin{equation}
    C = \dfrac{\Lambda\qty(1+\epsilon)-2\qty(\beta+3)\qty(1+\epsilon)\xi^2-3c^4\xi^6}{48\xi^4}\:.
\end{equation}
These changes result in the metric 
\begin{equation}
    ds^2  = \dfrac{dr^2}{B(r)} + B(r)e^{2\alpha(r)}\dfrac{r^2}{4}\qty(d\psi + \cos\theta d\varphi)^2 + \dfrac{r^2}{4}d\Omega^2\:,
\end{equation}
with functions 
\begin{equation}
    \alpha(r) = \dfrac{1}{2}\ln(\dfrac{1}{1-r^2\xi^2})\:,
\end{equation}
and 
\begin{equation}
    B(r) = \dfrac{r^2\xi^2-1}{6\xi^6r^4}\qty(\qty(6\xi^2 -\Lambda)\ln(\qty(r^2\xi^2-1)^2)-\Lambda\xi^2\qty(2r^2+\xi^2r^4)+12\xi^4r^2 + 6c^4\xi^6)\:.
\end{equation}
Looking at singularities and slight changes to this metric we may find two Eguchi-Hanson like spaces. We will analyze their singularities and GR limits.
\subsection{Physical properties}
The first property we examine will be the spacetime singularities. Calculation of the Ricci and Kretschmann invariants leads us to find 
\begin{equation}
    R = \dfrac{12\xi^2 + \Lambda\qty(6r^2\xi^2-8)}{3\qty(r^2\xi^2-1)}\:,
\end{equation}
and
\begin{equation}
    K = \dfrac{\frak{H}(r)}{\xi^{12} r^{12}\qty(r^2\xi^2-1)^2}\:.
\end{equation}
From these we see singularities at $r=1/\xi$ and $r=0$. It appears that $\xi=0$ is also a singularity, but in the limit that $\xi=0$ we find a finite result. We find for the $r$ range $0<r<1/\xi$ that we exhibit a 4D spatial solution, while when we pass this singularity and examine $1/\xi < r< \infty$ we are in the 3+1D Taub-NUT solution aforementioned. What we do notice is that only even powers of $\xi$ appear within our solution and so if we allow the change from $\xi$ to $\chi=i\xi$ we find a proper full 4D Eguchi-Hanson $f(R)$ solution with domain $0<r<\infty$.
Limit that $\xi=0$ or $\chi=0$ we find the Ricci become constant with the function $B(r)$ taking the standard $\Lambda$-Eguchi-Hanson
\begin{equation}
    B(r) = 1-\dfrac{c^4}{r^4}-\dfrac{\Lambda r^2}{9}\:.
\end{equation}
Given most do not want a cosmological constant in a 4D solution we set $\Lambda=0$ and find standard Eguchi-Hanson when $\chi=0$ or $\xi=0$ 

\section{Conclusion}\label{sec4}
Inspired by the existence of self-dual spaces in Einstein gravity, in this article, we find exact analytical solutions to the $f(R)$ gravity theory. The solutions are resembling the the Taub-NUT and Eguchi-Hanson geometries.  We consider a metric ansatz for the Taub-NUT solutions and find the action of the $f(R)$ gravity with an unknown $f(R)$ function. Varying the action, we find two classes of solutions for the Taub-NUT geometries.  To our knowledge, this is the first construction of the Taub-NUT solutions in $f(R)$ gravity with a coordinate dependant Ricci scalar. We analyze the physical behaviours of the obtained solutions, as well as their thermodynamics. We also find the orbits of the test particles around such solutions. Moreover, we find an Eguchi-Hanson solution and study its physical behaviours. 
 Moreover, in the context of the duality between the rotating black holes and the conformal field theory (CFT) in general relativity, we may find and look at the rotating solutions to the results of this article, to establish the existence (or non existence) of such duality for the rotating black hole solutions in $f(R)$ gravity. The duality has been shown to be valid through comparison between the macroscopic black hole quantities, as solutions of the general relativity, and the microscopic CFT quantities. In particular, in the context of duality, there is a perfect match between the macroscopic Bekenstein-Hawking entropy of the rotating black holes and the entropy of the CFT which is computed by the Cardy formula. Another very interesting result which supports the duality is coming from the study of the super-radiant scattering off the rotating black holes. It was shown that the bulk scattering amplitudes are in precise agreement with the scattering results from CFT. The scattering amplitudes of CFT are completely by the conformal invariance. Finding the higher dimensional Taub-NUT and Eguchi-Hanson solutions in $f(R)$ gravity is another interesting future work.
\section*{Acknowledgements}
The authors would like to thank Dr. Bardia H. Fahim for discussions. This work was supported by the Natural Sciences and Engineering Research Council of Canada.
\section*{References}
\bibliographystyle{unsrt}
\bibliography{refs}
\end{document}